\documentstyle[12pt]{article}
\def\lsim{\mathrel{\rlap {\raise.5ex\hbox{$ < $}}
{\lower.5ex\hbox{$\sim$}}}}

\newcommand{\pr}{\paragraph{}}
\newcommand{\be}{\begin{equation}}
\newcommand{\ee}{\end{equation}}
\newcommand{\bea}{\begin{eqnarray}}
\newcommand{\nn}{\nonumber}
\newcommand{\eea}{\end{eqnarray}}
\newcommand{\nd}[1]{/\hspace{-0.6em} #1}
\newcommand{\nk}{\noindent}
\def\gappeq{\mathrel{\rlap {\raise.5ex\hbox{$>$}}
{\lower.5ex\hbox{$\sim$}}}}
 
\def\lappeq{\mathrel{\rlap{\raise.5ex\hbox{$<$}}
{\lower.5ex\hbox{$\sim$}}}}
 
\begin{document}
 
\begin{titlepage}
\begin{flushright}
ACT-04/97 \\
CTP-TAMU-05/97 \\
quant-ph/9702003 \\
\end{flushright}

\begin{centering}
\vspace{.1in}
{\large {\bf
Microtubules: The neuronic system of the neurons?
}} \\
\vspace{.2in}
{\bf N.E. Mavromatos$^{a}$} 
and 
{\bf D.V. Nanopoulos$^{b,c,\diamond}$}

\vspace{.03in}
 
{\bf Abstract} \\
\vspace{.1in}
\end{centering}
{\small In this talk we review recent work 
on integrable models for Microtubule (MT) networks, subneural
paracrystalline cytosceletal structures,
which seem to play a fundamental role in the
neurons. We cast here the complicated
MT dynamics in the form of a $1+1$-dimensional
non-critical string theory, which can be 
formulated in terms of (dual) Dirichlet branes,
according to modern perspectives.
We suggest 
that the MTs are the microsites in
the brain, for the emergence of stable, macroscopic
quantum coherent states,
identifiable with the {\em preconscious states}.
Quantum space-time effects, as
described by non-critical string theory, trigger
then an {\em organized collapse}
of the coherent states
down to a specific or {\em conscious state}. The whole
process we estimate to
take ${\cal O}(1\,{\rm sec})$, in excellent agreement
with a plethora of experimental/observational
findings. 
The complete integrability of the stringy model for MT
proves sufficient
in providing a satisfactory solution
to memory coding and capacity. Such features
might turn out to be important
for a model of the brain as a quantum computer.}

\vspace{1in}

\vspace{1in}
\begin{flushleft}
$^\diamond$ Invited talk presented at the Workshop 
on ``Biophysics of Microtubules'', Texas Medical Center, 
Houston, Texas, April 1996. \\
$^{a}$ P.P.A.R.C. Advanced Fellow, Department of Physics
(Theoretical Physics), University of Oxford, 1 Keble Road,
Oxford OX1 3NP, U.K.  \\
$^{b}$ Department of Physics, 
Texas A \& M University, College Station, TX 77843-4242, USA, \\
$^{c}$ Astroparticle Physics Group, Houston, 
Advanced Research Center (HARC), The Mitchell Campus,
Woodlands, TX 77381, USA, \\

\end{flushleft}

\end{titlepage} 
 
\newpage
\pr
The interior of living cells is structurally
and dynamically organized by {\it cytoskeletons}, i.e.
networks of protein polymers. Of these structures,
{\it MicroTubules} (MT) appear to be \cite{hameroff}
the most fundamental. 
These are paracrystalline cytoskeletal
structures which seem to play a fundamental r\^ole 
for the cell {\it mitosis}, as well as 
for the transfer of electric signals
and, more general, for dissipation-free energy 
transfer in the cell, according to ideas of 
Fr\"ohlich (1986)~\cite{Frohlich}. 
MT networks are, therefore, also important
for neuronic cells. In this latter
respect, according to recent ideas~\cite{HP,ideas}, 
neuronic MT are believed to be associated 
with conscious perception.
This conjecture was based on the 
observation that certain micro-organisms
without a nervous system
but with cytoskeletal structure, 
c.f. {\it paramitium}, 
exhibit
some sort of unexpected awareness.
Some 
scenaria have been developed in order to provide 
physical mechanisms for such a phenomenon~\cite{ideas}.
According to these, the MT networks may  sustain 
{\it macroscopic} 
coherent quantum mechanical states,
identfied with the {\it preconscious states}. 
Coupling the latter 
to space-time quantum (fluctuating) gravitational 
degrees of freedom  
triggers an {\it organized collapse} down to 
a specific or {\it conscious} state.
Such daring assumptions/conjectures
have gain some support by the fact that in 
certain physical models of MT dynamics
the estimated time of collapse 
is of order ${\cal O}(1~{\rm sec})$, which is 
in excellent qualitative agreement 
with a plethora of experimental/observational
findings in Neurobiology. 
\pr
In ref. \cite{HP}, however, no specific 
physical model has been constructed, which could
constitute a physical realization of the above-described
phenomenon. 
Such an attempt was made
in ref. \cite{mn}, where 
there was presented
a microscopic model for 
MT quantum dynamics, realizing the above scenario,
yielding a collapse time 
of ${\cal O}(1~{\rm sec})$. It is the purpose of this
talk to review the physical aspects of that work,
and also to cast it in a more recent perspective
following subsequent developments in the field 
of stringy quantum gravity, which occured over the 
past year. 
\pr
First, let us recapitulate the experimental findings
concerning 
MT that will be useful in our analysis. 
MT are hollow cylinders 
comprised of an exterior surface
(of cross-section
diameter
$25~nm$)
with 13 arrays
(protofilaments)
of protein
dimers
called tubulines.
The interior of the cylinder
(of cross-section
diameter $14~nm$)
contains ordered water molecules,
which implies the existence
of an electric dipole moment and an electric field.
The arrangement of the dimers is such that, if one ignores
their size,
they resemble
triangular lattices on the MT surface. Each dimer
consists of two hydrophobic protein pockets, and
has an unpaired electron.
There are two possible positions
of the electron, called $\alpha$ and $\beta$ {\it conformations}. 
When the electron is
in the $\beta$-conformation there is a $29^o$ distortion
of the electric dipole moment as compared to the $\alpha $ conformation.
\pr
In standard models for the simulation of the MT dynamics,
the `physical' degree of freedom -
relevant for the description of the energy transfer -
is the projection of the electric dipole moment on the
longitudinal symmetry axis (x-axis) of the MT cylinder.
The $29^o$ distortion of the $\beta$-conformation
leads to a displacement $u_n$ along the $x$-axis,
which is thus the relevant physical degree of freedom.
This way, the effective system is one-dimensional (spatial),
and one has a first indication that quantum integrability
might appear. We shall argue  later on
that this is indeed the case.
\pr
Information processing
occurs via interactions among the MT protofilament chains.
The system may be considered as similar to a model of
interacting Ising chains on a trinagular lattice, the latter being
defined on the plane stemming from fileting open and flatening
the cylindrical surface of MT.
Classically, the various dimers can occur in either $\alpha$
or $\beta$ conformations. Each dimer is influenced by the neighboring
dimers resulting in the possibility of a transition. This is
the basis for classical information processing, which constitutes
the picture of a (classical) cellular automatum.
\pr
The quantum computer character of the MT network results
from the assumption that each dimer finds itself in a
superposition of $\alpha$ and $\beta$ conformations \cite{HP}.
There is a macroscopic
coherent state among the various chains, which
lasts for ${\cal O}(1\,{\rm sec})$ and constitutes the `preconscious'
state~\cite{mn}. The interaction
of the chains with (stringy) quantum gravity, then, induces
self-collapse of the wave function of the coherent MT network,
resulting in quantum computation.
\pr
In ref. \cite{mn} we assumed, 
for simplicity,  that the collapse occurs
mainly due to the interaction of each chain with
quantum gravity, the interaction from neighboring chains
being taken into account by including mean-field interaction terms
in the dynamics of the displacement field of each chain. This amounts
to a modification of the effective potential by
anharmonic oscillator terms.
Thus, the effective system under study is two-dimensional,
possesing one space and one time coordinate. The precise
meaning of `time' in our model will be clarified when
we discuss the `non-critrical string' representation
of our system.
\pr
Let $u_n$ be the displacement field of the $n$-th dimer in a MT
chain.
The continuous approximation proves sufficient for the study of
phenomena associated with energy transfer in biological cells,
and this implies that one can make the replacement
\be
  u_n \rightarrow u(x,t)
\label{three}
\ee
with $x$ a spatial coordinate along the longitudinal
symmetry axis of the MT. There is a time variable $t$
due to
fluctuations of the displacements $u(x)$ as a result of the
dipole oscillations in the dimers.
At this stage, $t$ is viewed as a reversible variable.
The effects of the neighboring
dimers (including neighboring chains)
can be phenomenologically accounted for by an effective
double-well potential \cite{mtmodel}
\be
U(u) = -\frac{1}{2}A u^2(x,t) + \frac{1}{4}Bu^4(x,t)
\label{four}
\ee
with $B > 0$. The parameter $A$ is temperature
dependent. The
model of ferroelectric distortive spin chains
of ref. \cite{collins} can be used to
give a temperature
dependence
\be
   A =-|const|(T-T_c)
\label{temper}
\ee
where $T_c$ is a critical temperature of the system, and
the constant is determined phenomenologically \cite{mtmodel}.
In realistic cases the temperature $T$ is very close to $T_c$,
which for the human brain is taken to be the room temperature
$T_c = 300K$.
Thus, below $T_c$
$A > 0$.
The important relative minus sign in the potential (\ref{four}), then,
guarantees the necessary degeneracy, which is necessary for the
existence of
classical solitonic solutions. These constitute the basis
for our coherent-state description of the
preconscious state.
\pr
The effects of the surrounding water molecules can be
summarized by a viscuous force term that damps out the
dimer oscillations,
\be
 F=-\gamma \partial _t u
\label{six}
\ee
with $\gamma$ determined phenomenologically at this stage.
This friction should be viewed as an environmental effect, which
however does not lead to energy dissipation, as a result of the
non-trivial
solitonic structure of the
ground-state
and the non-zero constant
force due to the electric field.
This is a well known result, directly relevant to
energy transfer in biological systems \cite{lal}.
\pr
To determine a dynamical equation of motion 
for the MT dimers, each having a mass $M$,
one should 
include a phenomenological kinetic term 
in the respective equation.  
It is also convenient~\cite{mtmodel,mn}
to use
a normalized
displacement field
\be
  \psi (\xi ) = \frac{u (\xi)}{\sqrt{A/B}}
\label{eight}
\ee
where,
\be
               \xi \equiv \alpha( x - v t)
\qquad \alpha \equiv \sqrt{\frac{|A|}{M(v_0^2 - v^2)}}
\label{nine}
\ee
with
\be
v_0 \equiv \sqrt{k/M} R_0
\label{sound}
\ee
the
sound velocity, of order $1 km/sec$,
and $v$ the propagation velocity
to be determined below. 
Above $k$ is a stiffness parameter, $R_0$ is the equilibrium
spacing between adjacent dimers. 
In terms of the $\psi (\xi )$ variable
the equation of motion 
acquires the form of the equation
of motion of an anharmonic oscillator in a frictional environment
\bea
 \psi '' &+& \rho \psi ' - \psi ^3 + \psi + \sigma = 0  \nn \\
\rho &\equiv& \gamma v [M |A|(v_0^2 - v^2)]^{-\frac{1}{2}}, \qquad
\sigma = q \sqrt{B}|A|^{-3/2}E
\label{ten}
\eea
where the prime denotes differentiation with respect to $\xi$, 
and $E$ is the electric field
due to the `giant dipole' representation of the MT cylinder,
as suggested by the experimental results \cite{mtmodel},
and $q=18 \times 2e$ ($e$ the electron charge) is a
mobile charge. 

Equation (\ref{ten}) has a {\it unique} bounded solution \cite{mtmodel}
\be
    \psi (\xi ) = a + \frac{b -a}{1 + e^{\frac{b-a}{\sqrt{2}}\xi}}
\label{eleven}
\ee
with the parameters $b,a$ and $d$ satisfying:
\be
(\psi -a )(\psi -b )(\psi -d)=\psi ^3 - \psi -
\left(\frac{q \sqrt{B} }{|A|^{3/2}} E\right)
\label{twelve}
\ee
According to ref. \cite{lal} the importance
of the force term $ qE $ lies in the fact
that eq (\ref{ten}) admits displaced classical soliton solutions
(kinks) with no energy loss.
The kink propagates along the protofilament axis
with fixed velocity
\be
    v=v_0 [1 + \frac{2\gamma^2}{9d^2Mv_0^2}]^{-\frac{1}{2}}
\label{13}
\ee
This velocity depends on the strength of the electric
field $E$ through the dependence of $d$ on $E$ via (\ref{twelve}).
Notice that, due to friction, $v \ne v_0$, and this is essential
for a non-trivial second derivative term in (\ref{ten}), necessary
for wave propagation.
For realistic biological systems $v \simeq 2 m/sec$.
With a velocity of this order,
the travelling waves
of kink-like excitations of the displacment field
$u(\xi )$ transfer energy
along a moderately long microtubule
of length $L =10^{-6} m$ in about
\be
t_T = 5 \times 10^{-7} sec
\label{transfer}
\ee
This time is very close to Frohlich's time for
coherent phonons in biological system~\cite{Frohlich}.
We shall come back to this issue later on.
\pr
The total energy of the solution (\ref{eleven}) is
easily calculated to be \cite{mtmodel}
\be
   E = \frac{1}{R_0} \int _{-\infty}^{+\infty} dx H
= \frac{2\sqrt{2}}{3}\frac{A^2}{B} +  \frac{\sqrt{2}}{3}
k \frac{A}{B} + \frac{1}{2} M^{*} v^2  \equiv \Delta + \frac{1}{2}
M^{*} v^2
\label{energy}
\ee
which is {\em conserved} in time.
The `effective' mass $M^{*}$ of the kink is given
by
\be
            M^{*} = \frac{4}{3\sqrt{2}}\frac{MA\alpha }{R_0 B}
\label{effmass}
\ee
The first term of equation (\ref{energy})
expresses the binding energy of the kink
and the second the resonant transfer energy.
In realistic
biological models the sum of these two terms
dominate over the third
term, being of order of $1eV$ \cite{mtmodel}. On the other hand,
the effective mass in (\ref{effmass}) is\cite{mtmodel} of order
$5 \times 10^{-27} kg$, which is about
the proton mass ($1 GeV$) (!).
As
we discussed in ref. \cite{mn}, and 
shall discuss later on, these values are essential
in yielding the correct estimates for the
time of collapse of the `{\it preconscious}' state due to our
quantum gravity environmental entangling. 
\pr
To make plausible
a consistent
study of such effects, one has to 
represent 
the equation of motion (\ref{ten})
as being derived from  string theory.
In such a framework, and in particular 
in a non-critical string context~\cite{aben,DDK},
the authors together with John Ellis~\cite{emn}
had already developed a theory of information storage
in quantum space-times with (black-hole) 
singulartities. The theory implies an  
irreversible time evolution for the matter system 
in interaction with the gravitational environment 
in a quite natural 
and mathematically rigorous way. 
The purpose of ref. \cite{mn} was to apply these ideas
to the dynamics of brain MT in an attempt to 
present a semi-microscopic model for quantum brain 
function in the spirit of refs. \cite{HP,ideas}.
\pr
For this purpose,
it is important to notice that the relative sign (+)
between the second derivative and the linear term in $\psi $
in
equation (\ref{ten}) is such that this equation
can be considered as corresponding to the
tachyon $\beta$-function equation
of
a (1 + 1)-dimensional string theory, in a
flat space-time with a dilaton field $\Phi $
linear in the {\it space-like}
coordinate $\xi $ \cite{aben},
\be
        \Phi = - \rho \xi
\label{dilaton}
\ee
Indeed, the most general form of a `tachyon' deformation
in such a string theory, compatible with
conformal invariance is that of a travelling wave \cite{polch}
$T (x ' )$ , with
\bea
  x'=\gamma _{v_s} (x - v_s t)
  \qquad &;& \qquad t'= \gamma _{v_s}(t-v_s x) \nn \\
\gamma _{v_s} \equiv (1 &-& v_s^2)^{-1/2}
\label{st7}
\eea
where $v_s$ is the propagation velocity of the string `tachyon'
background.
As argued in ref. \cite{polch} these translational invariant
configurations
are the most general
backgrounds,
consistent
with a {\it unique}
factorisation of the string $\sigma$-model
theory on a Minkowski space-time $G_{\mu\nu} = \eta _{\mu\nu}$
\be
  S= \frac{1}{4\pi \alpha '} \int d^2 z
  \left[\partial X^{\mu} {\overline \partial } X^{\nu}
G_{\mu\nu}(X) + \Phi(X) R^{(2)} + T(X)\right]
\label{st2b}
\ee
into two conformal field theories,
for the $t'$ and $x'$ fields,
corresponding to central charges
\be
c_{t'} = 1 -24 v_s^2 \gamma _{v_s}^2  \qquad ; \qquad
c_{x'} = 1 + 24 \gamma _{v_s}^2
\label{st8}
\ee
In our case (\ref{ten}), the r\^ole of the space-like
coordinate $x'$
is played by $\xi $ (\ref{nine}).
The velocity of light in this effective string
model is replaced by the sound velocity $v_0$ (\ref{sound}),
and the velocity $v_s$
is defined in terms of
the velocity $v$
of the kink (\ref{13}) by
expressing~\cite{polch}
the friction coefficient
$\rho $ in terms of the central charge deficit
(\ref{st8}) 
\be
  \rho = \sqrt{\frac{1}{6}(c(\xi) - 1)} = 2 \gamma _{v_s}
\label{qdef}
\ee
The space-like `boosted' coordinate $\xi$, thus, plays the r\^ole
of space in this effective/Liouville mode string theory.
\pr
Notice that with the definition (\ref{qdef}) the local-field-theory
kink solution (\ref{eleven}),
propagating with a {\it real} velocity $v$,
is mapped to a non-critical string background which propagates
with a velocity $v_s$ that could be {\it imaginary} ($v_s^2 < 0$).
Indeed,
the condition for reality of $v_s$ can be easily found
from (\ref{ten},\ref{13},\ref{qdef}) to be
\be
      8M|A| < 9d^2Mv_0^2
\label{reality}
\ee
It can be easily seen that
for the generic values of the parameters of the MT model
described above~\cite{mtmodel},
(\ref{reality}) is not satisfied, which implies that, when
formulated as a
non-critical string theory, the MT system corresponds
to a $1 + 1$ dimensional non-critical string
with Wick-rotated `time' $s=it$~\cite{polch},
and, therefore, corresponds to a matter central charge
that overcomes the $c=1$ barrier
\be
25 \ge c_{s'} = 1 + 24 |v_s|^2 (1 + |v_s|^2 )^{-1} \ge 1
\label{barier}
\ee
In this regime, Liouville theory is poorly understood, but
there is the belief that this range of matter central charges
is characterized by
polymerization properties
of random surfaces. From our point of view this may be related
to certain growth properties of the MT networks, which are
discussed in ref. \cite{mn}.
\pr
An interesting question arises as to whether there are
circumstances under which (\ref{reality}) is satisfied,
in which case the matter content of the theory
is characterized by the central charge $c_{t'} \le 1 $ (\ref{st8}).
To this end, we note that
the parameters with the largest uncertainty
in the relation (\ref{reality})
are $A$ and $d$.
At present,
there are no accurate experimental data for
$A$, which depends on temperature (\ref{temper}).
The parameter $d$ is sensitive
to the order of magnitude of the electric field $E$.
The latter is non-uniform along the MT axis. There is a
sharp increase of $E$ towards the end points
of the MT protofilament axis ~\cite{mtmodel},
and for such large $E$, $d \sim E^{1/3}$.
Due to
(\ref{13}), an increase in $d$ results in
an increase in the kink velocity $v$
at the end points of the MT.
For realistic
biological systems, under normal circumstances,
the increase in $v$
can be even up to two orders of magnitude, resulting
in kink velocities
of
${\cal O}[10^2 m/sec]$~\cite{mtmodel}
at the end points of the MT.
\pr
As we shall argue now,
both parameters $A$ and $d$
can be drastically affected
by an abrupt distortion of the
environment due to the influence of an
{\it external stimulus}. For instance,
it may be possible for
such abrupt distortions
to cause a local disturbance among the dimers,
so that
the value of $A$ is momentarily
diminished significantly and the
condition
(\ref{reality})
is met.
However, such a distortion might affect the
order of magnitude estimates of the
effective mass scales involved in the problem, as we
discussed above (\ref{effmass}), and this may have consequences
for the decoherence time. So, we shall not consider it
for the purposes of this work.
More plausibly, an abrupt
environmental
distortion
will lead to a sudden increase in the
electric field $E$, which, in turn,
results in the
formation of a fast kink of $v \sim v_0$, via an increase in
$d$.
As we have seen above, this is not
unreasonable for the range of the parameters
pertaining to realistic biological systems.
In such cases the (fast) kinks are represented as
translational-invariant backgrounds of
non-critical string theories,
propagating with {\it real} velocities $v_s$ (\ref{qdef}).
This is the kind of structures that we shall mostly be
interested in for the purposes of this work.
Coupling them
to  quantum gravity will
lead to the collapse of the preconscious state,
as we shall discuss in the next
section.
\pr
The important advantage of formulating the MT system as
a $c=1$ string theory, lies in the possibility
of casting the friction problem in a Hamiltonian
form~\cite{emn,mn}. 
This is quite important for the 
{\it canonical} quantization of the kink solution,
which will provide us with a concrete example of
a large-scale quantum coherent state for the preconscious
state~\cite{mn}. An explicit construction
of such quantum solitonic coherent 
states has been given in ref. 
\cite{mn}. 
Such
large-scale coherent states
in the MT networks may, thus, be
considered responsible for loss-free energy
transfer along the tubulines.
\pr
Suppose now that {\it external stimuli}
produce  sufficient distortion in the
electric dipole moments of the water environment
of the MT.
As a result,
conformational (quantum) transitions
of the tubulin
dimers occur.
Such abrupt pulses
may cause sufficient distortion of the
space time
surrounding the tubulin dimers,
which in turn
leads to the formation of {\it virtual}
`black holes' in the effective target two-dimensional
space time.
Formally this is expressed by coupling the
$c=1$ string theory to two-dimensional quantum gravity~\cite{mn}.
This elevates the matter-gravity system to a
critical $c=26$ theory.
Such a coupling, then,
causes decoherence, due to induced instabilities
of the kink quantum-coherent `preconscious state',
in a way described in ref. \cite{mn}.
As
the required
collapse time of ${\cal O}(1\,{\rm sec})$ of
the wave function of the coherent
MT network is several orders of magnitude bigger than the
energy transfer
time  $t_T$ (\ref{transfer}),
the two mechanisms are compatible with each other.
Energy is transfered during the quantum-coherent
preconscious state, in $10^{-7} sec$,
and then collapse occurs to a certain (classical) conformational
configuration. In this way, Frohlich's frequency~\cite{Frohlich}
associated with coherent `phonons' in biological
cells is recovered, but in a rather different setting.
\pr
Once a virtual black hole is formed in a MT chain,
the subsystem
of the displacement modes $\psi (\xi)$ becomes {\it open}
in a statistical mechanics sense.
This subsystem is a (1+1)-dimensional
non-critical (Liouville) theory. It is known that
the singularity structure of black holes in such
systems is described by
a topological $\sigma$-model,
obtained by twisting the $N=2$ supersymemtric
black hole~\cite{wbreak}. The field theory {\it
at the
singularity} is described by an enhanced
topological $W_{1+\infty} \otimes W_{1+\infty}$
symmetry. {\it Away} form the singularity,
such symmetry is {\it broken spontaneously }
down to a single $W_{1+\infty}$ symmetry,
as a result of the non-vanishing target-space gravitational
condensate~\cite{wbreak}. Spontaneous breaking in a
(1+1)-dimensional target string theory is allowed, in the sense that
the usual infrared infinities that prevented it
from happening in a local field theory setting are absent.
In ref. \cite{wbreak,emn} we have demonstrated
this phenomenon explicitly
by showing that, due to the (twisted)
$N=2$ supersymmetry associated with the topological
nature of the singularity, there is a suppression of the
tunneling effects, which in point-like theories
would prevent the phenomenon from occuring.
As a result,
the appearance of {\it massless}
states takes place. 
Such states are delocalized global states, belonging to
the lowest-level of the string spectrum~\cite{wbreak}.
These are the leg-poles that appear in the scattering
amplitudes of $c=1$ Liouville theory~\cite{legpoles}.
Their excitation in the brain results in
conscious perception, in a way similar to
the one argued in the context of local field theory
regarding
the excitation of the dipole wave quanta~\cite{delgiud}.
Here, however, 
the mechanism of conscious perception
appears formally much more complicated, due to the
complicated nature of the enormous
stringy symmetries that are spontaneously
broken in this case.
\pr
{}From a formal point of view, the formation of virtual black holes,
with varying mass, can be modelled by the action of
world-sheet instanton deformations~\cite{emn}. The latter
have the property of shifting (renormalizing)  the
Wess-Zumino level parameter $k$ (related to central charge)
of the black hole conformal field theory of ref. \cite{witt}.
From a conformal field theory point of view,
instantons are associated with induced {\it extra logarithmic}
divergencies (on the world sheet) in the presence of the
matter leg-poles. In our approach to target time, $t$, as a dynamical
world-sheet
renormalization group scale~\cite{emn}, $\phi = -t$,
such logarithmic divergencies, when
regularized, lead to extra
time dependences in the central charge of the theory, and
hence to a
time-dependent `effective' $k$, as mentioned above.
It can be shown~\cite{emn} that
in such a case the $ADM$ mass of the black hole~\cite{adm}
depends on the scale (time)
\be
  M_{bh} \propto \frac{1}{\sqrt{k(t)-2}} e^a
\label{adm}
\ee
where $a$ is a constant that can be added to the dilaton field
without affecting the conformal invariance of the black hole solution
without matter. As shown in \cite{emn}, $k(t)$ actually increases
with time $t$, leading asymptotically to $\infty$, which corresponds
to the flat space-time limit. In that case, the system keeps `memory'
of the dilaton constant $a$, which pre-existed the black hole
formation. From a MT point of view, the constant $a$ corresponds
to a spontaneously chosen vacuum string state as a result
of spontaneous breaking mechanisms of, say, electric dipole
rotational symmetries etc,
in the ordered water
environment~\cite{delgiud}.
This is important
for {\it coding} of memory states in this framework.
From the above discussion it becomes clear
that {\it memory} operation in our approach is a two step
process: (i) formation of a black hole, of a fixed value
for the dilaton vacuum expectation value (vev)
$a$, related to spontaneous
breaking occuring in the water environment, as a result
of an {\it external stimulus}, and (ii) evaporation of
the black hole, due to {\it quantum} instabilities,
described formally by world sheet instanton effects in our
completely integrable approach to MT dynamics.
The latter effects drive the black hole
to a vanishing-mass limit by shifting $k$ and
{\it not } the dilaton vev, $a$. This implies
{\it storage} of information, according to
the general ideas of ref. \cite{emn}.
Indeed,
from a conformal field theory point of view, the constant $a$
can be shifted by exactly marginal deformations
(moduli) of the black hole $\sigma$ model, whose couplings
are {\it arbitrary}.
Such an operator has been constructed explicitly in ref.
\cite{chaudh},
and consists exclusively of combinations of global state deformations.
It 
constitutes one of the (infinite number of) $W_\infty$ charges
that have been conjectured to characterize a two-dimensional
stringy black hole~\cite{emn}.
\pr
We now consider the propagation of low-energy matter modes, $T(\xi)$, 
in such black hole space times. 
From the point of view of $(1+1)$-dimensional string theory
such modes correspond to lowest-lying  string states, known as `tachyons'.
In our stringy representation of MT such excitations correspond to
the  displacement field $\psi (\xi)$~\cite{mn}. 
The important point to notice is that the
system of $T(\xi )$ coupled to a black hole
space-time,
even if the latter is a virtual configuration, 
cannot be critical
(conformal invariant) {\it non-perturbatively}
if the tachyon has a
travelling wave form. The factorisation
property of the world-sheet action  (\ref{st2b})
in the flat target space-time case breaks down due to the
non-trivial target-space graviton structure.
Then a travelling wave cannot be compatible with
conformal invariance, and renormalization scale dressing
appears necessary.
The gravity-matter system is viewed as a $c=26$ string
\cite{witt,emn}, and hence the
renormalization scale is {\it time-like} \cite{aben}.
This implies time dependence in $T (\xi, t)$.
\pr
A natural question arises whether there exist a deformation
that turns on the coupling $T (\xi )$ which is exactly
marginal so as to maintain conformal invariance.
The exaclty marginal
deformation of this black hole background
that turns on matter, $L_0^1{\overline L}_0^1$
in the notation of ref. \cite{chaudh},
couples necessarily
the propagating tachyon $T(\xi )$ zero modes to
an infinity of higher-level string states \cite{chaudh}.
The latter are classified according to
discrete representations
of the $SL(2,R)$ isospin, and together with the
propagating modes,
form
a target-space $W_\infty$-algebra \cite{emn,bakas}.
This coupling of massive and massless modes
is due to the non-vanishing Operator Product Expansion
(O.P.E.)
among the vertex operators of the
$SL(2,R)/U(1)$ theory \cite{chaudh}. This theory 
possesses
an infinity of conserved charges in target space
\cite{emn} corresponding to the Cartan subalgebra
of the infinite-dimensional $W_\infty$ \cite{bakas}.
In practice, such global charges, which
contribute phase factors to the string universe wave function,
are impossible to measure by localized scattering
experiments in our world.
This, as explained in \cite{emn},
leads to the effective
breakdown of quantum coherence in the low-energy world~\cite{emn}.
From the point of view of MT, such $W$ modes might be thought of as
constituting the
{\it `consciousness degrees of freedom'}~\cite{mn},
which in this picture, are not exotic, as suggested in ref. \cite{Page},
but exist already in a string
formalism, and they result in the {\it complete integrability }
of the two-dimensional black hole Wess-Zumino model.
\pr
This integrability persists quantization \cite{wu},
and it is very important for the quantum coherence
of the string black hole space-time\cite{emn}. Due to the
specific nature of the $W_\infty$ symmetries, there
is no information loss during a stringy black hole decay,
the latter being
associated
with
instabilities
induced by higher-genus effects on the world-sheet
\cite{emndec,emn}. The phase-space volume of the
effective field theory is preserved in time, {\it only
if} the infinite set of the global string modes
is taken into account. This is due to the string-level mixing
property of the $W_\infty$ - symmetries of the target space.
\pr
However,
any local operation of measurement, based on local scattering
of propagating matter, such as the functions performed
by the human brain, will necessarily break this coherence,
due to the truncation of the string deformation spectrum
to the localized propagating modes $T(\xi)$. The latter will, then,
constitute a subsystem
in interaction with an
{\it environment} of global string modes.
The quantum integrability of the full string system is crucial in
providing the necessary couplings.
This breaking of coherence
results in an arrow of time/Liouville scale,
in the way
explained briefly
above \cite{emn}. The black-hole $\sigma$-model
is viewed as a $c=26$ critical string, while the travelling
wave background is a non-conformal deformation.
To restore criticality one has to
dress $T(\xi )$ with a Liouville time dependence
$T(\xi, t)$ \cite{emn}.
We stress, once more, that
the Liouville renormalization scale now is time-like,
in contrast with the previous string picture
of a $c=1$ matter string theory, representing the
displacement field $\psi $ alone before coupling to gravity.
In this framework, one can 
derive~\cite{emn} an equation for the temporal
evolution of the density matrix 
$\rho $ of the subsystem 
comprising of the 
Liouville-dressed displacement field $T(\xi, t)$
in interaction with the fluctuating space-time.
The equation is of the generic form 
expected in an effective theory of quantum 
gravity~\cite{ehns}:
\be
\partial _t \rho = i[\rho, H] + \nd{\delta H}\rho
\label{modlio}
\ee
where $\nd{\delta H}$ is a non-commutator term,
depending on data of the (non-critical) $\sigma$-model 
that describes propagation of the string excitations 
in a non-trivial black-hole space time~\cite{emn,mn}.    
\pr
One can calculate in this approach
the off-diagonal
elements of the density matrix in the string theory space $u^i$,
with now $u^i(t)$ representing the displacement field of the
$i$-th dimer~\cite{mn}. 
The computation proceeds analogously \cite{emn}
to the
Feynman-Vernon \cite{vernon}  and
Caldeira and Leggett \cite{cald}
model of environmental oscillators, using the influence functional
method, generalized properly to the string theory
space
$u_i$. The
general theory of time as a world-sheet
scale predicts \cite{emn}
the following
order-of-magnitude estimate for the decoherence time:
\be
    t_{col} =O[\frac{M_{gus}}{E^2N}]
\label{colltime}
\ee
where $E$ is  a typical energy scale in the problem,
and $M_{gus}$ is a string scale, characterising the 
dynamics of the space-time environment of the 
effective MT model. 
If we take $M_{gus}$ to be the realistic four-dimensional 
quantum string gravity scale $10^{18}$ GeV~\cite{mn}, the 
we can 
estimate
that a collapse time 
of ${\cal O}(1\,{\rm sec})$ is compatible with a number of coherent
tubulins of order
\be
 N \simeq 10^{12}
\label{number}
\ee
provided that the
energy stored in the kink background is of the
order of
$eV$. This
is indeed
the case of the (dominant) sum of binding  and
resonant transfer energies $\Delta \simeq 1 eV $ (\ref{energy})
at room temperature
in the
phenomenological model of ref. \cite{mtmodel}.
This number of tubulin dimers
corresponds to a fraction of $10^{-7}$ of the total
brain,
which is pretty close to the fraction believed
to be responsible for human perception on the basis of
completely independent biological methods.
\pr
Having described the basics of our mechanism 
for conscious perception, we went on in ref. \cite{mn}
to discuss issues related to memory {\it coding}
and {\it capacity} of the brain as a quantum 
computer. 
For this purpose one should first stress
the importance of the surrounding water molecules
for the proper functioning or even the existence
of MT~\cite{delgiud}. As a result of
its electric dipole structure,
the ordered water environment 
has been conjectured to exhibit a laser-like
behaviour in ref.~\cite{prep}. In local field theory 
models of the dynamics of MT~\cite{umez,stewart,delgiud}
coherent modes emerge as a result of the
interaction of the
electric dipole moments
of the water molecules
with the quantized electromagnetic radiation.
Such quanta can be understood~\cite{delgiud}
as Goldstone modes arising from the spontaneous
breaking of the electric dipole symmetry, which in the
work of ref. \cite{delgiud} was the only symmetry
to be assumed spontaneously broken.
In our string model,
as discuss
in \cite{mn}, and mentioned above
a more complicated (infinite-dimensional) $W_\infty \otimes 
W_\infty $ symmetry
breaks spontaneously, which incorporates the simple rotational
symmetry of the point-like theory models.
The emergence of coherent dipole quanta resembles
the picture of Fr\"ohlich
coherent `phonons'~\cite{Frohlich},
emerging in biological
systems for energy transfer without dissipation.
\pr
In our non-critical-string approach
the existence of such coherent states in the surrounding
water results in the friction term proportional to $\rho$
in (\ref{ten}). What we have argued above is that,
because of this interaction,
a kink soliton can be formed, provided that
the MT are of sufficient length. Such solitons
can then be themselves squeezed coherent states, being responsible
for a  {\it preconscious} state of the mind. 
In local field theory models of the brain {\it external stimuli}
are believed responsible for triggering
spontaneous breaking of the electric dipole rotational
symmetry of the water environment.
The collective Goldstone modes of such a
breaking (dipole quanta)
are spin wave quanta and the
system's phases are macroscopically characterized
by the value of the order parameter, which in this case
is the polarization ({\it coding}). If the
ground state of the system is considered as the {\it memory}
state, then the above process is just {\it memory printing}.
In this picture,
{\it memory recall} corresponds to the excitation,
under another external stimulus,
of dipole quanta of similar nature to those leading
to the printing.
The brain, then,
`consciously feels'~\cite{stewart,vitiello} the pre-existing order
in the ground state.
\pr
The main problem encountered in most of the local field
theory models of the brain is related to memory capacity,
as explained in \cite{mn}, since those models contain only 
a {\it finite} amount of conserved quantum numbers, 
and therefore are insufficient to store the enormous 
amount of information believed to be stored by the human brain.
An attempt to overcome those difficulties has been made in
the interesting work of ref.
\cite{vitiello}, making 
use of
{\it dissipative} models for brain function, within a local field theory
framework. As observed in
ref. \cite{vitiello}, the doubling of degrees of freedom
which appears necessary for a canonical quantization
of an open system in a dissipative environment~\cite{umez,umeztfd},
is essential in yielding~\cite{harm} a {\it non-compact}
$SU(1,1)$ symmetry for the system of damped harmonic oscillators,
used as a toy example for simulating quantum brain physics.
The quantum numbers of such a system are the $SU(1,1)$ isospin
and its third component, $j \in Z_{\frac{1}{2}}, m \ge |j|$.
The memory (ground) state corresponds to $j=0$ and there is a huge
degeneracy  characterised by the various coexisting (infinite)
eigenestates
of the Casimir operator  for the $SU(1,1)$ isospin.
The open-character of the system introduces a time arrow
which is associated with the {\it memory printing} process and
is compatible with the `observation' that
`only the past can be recalled'~\cite{vitiello}.
As far as we can see,
the problem with this approach
is that it necessarily introduces
dissipation in the energy functional, through the non-hermitian
terms in the interaction hamiltonian between the subsystem and the
environment~\cite{umez,vitiello}.
Hence, it is not easy to see how to reconcile this
with the above-mentioned property of biological systems to
transfer energy without
dissipation across the cells~\cite{Frohlich,lal}.
Moreover, from our point of view, this approach cannot
take into account realistic quantum gravity effects,
which according to the hypothesis 
of refs. \cite{HP,ideas}
are considered responsible for conscious perception.
\pr
Our Stringy 
approach to the 
MT dynamics~\cite{mn} seems to provide a way out of these problems~\cite{emn}
due to the infinite-dimensional
gauge stringy symmetries that mix the various levels.
In the (completely integrable)
black hole model of ref. \cite{witt}, which is used to simulate
the physics of the MT~\cite{mtmodel}, there is an undelrying world-sheet
$SL(2,R)$
symmetry of the $\sigma$-model,
according to which the various stringy states are classified.
The various states of the model, including global string modes
characterised by discrete values of (target) energy and momenta,
are classified by the non-compact isospin $j$ and its
third component $m$, which - unlike the compact
isospin $SU(2)$ case - is not restricted by the  value of $j$.
Thus, for a given $j$, which in the case of string states plays the role
of energy, one can have an {\it infinity} of states labelled by the
value of the third isospin component $m$.
All such states are characterised by a $W_{1+\infty}$ symmetry
in target space. As we mentioned above, this symmetry
is responsible for the maintenance of
{\it quantum coherence} in the presence of a
black hole background~\cite{emn}, in the sense of an area-preserving
diffeomorphism in a matter phase-space of the two-dimensional target
space theory.
Thus, 
for each matter ground state of a propagating degree of freedom,
say the zero mode of the massless field corresponding to the static
``tachyon'' background of ref. \cite{witt}, with $SL(2,R)$
quantum numbers
$j=-\frac{1}{2}, m=0$, there will be an infinite (energy) degeneracy
corresponding to a continuum of black hole space-time backgrounds
with different $ADM$ masses.
These backgrounds are
essentially generated by
adding various constants to the configuration of the dilaton
field in this
two-dimensional string theory~\cite{witt,chaudh}.
It should be noted here that the infinity of
propagating
``tachyon'' states (lowest string mass-level (massless) states),
corresponding to other values of $m$, for
continuous representations of $j$, constitute
{\it excitations} about the ground state(s), and, thus,
they should not be considered as contributing to the
ground state degeneracy.
In principle, there may be an additional infinity of
quantum numbers corresponding to higher-level $W$-hair charges
of the black hole space time which are
believed~\cite{emn}
responsible for quantum coherence at the full string theory level.
\pr 
Taking into account the conjecture of the present work,
that formation of virtual black holes
can occur in brain MT models, which
would correspond to different modes of collapse of
pulses of the displacement field $\psi $
defined in (\ref{eight}),
one obtains a system of {\it coding} that is capable to
solve in principle the problem of memory capacity.
Information is stored in the brain in the following sense:
every time there is an external stimulus that brings the brain
out of equilibrium, one can imagine an abrupt conformational
change of the MT dimers, leading to a collapse
of the pulse pertaining to the displacement field. Then a (virtual)
black hole
is formed leading to a spontaneous collapse of the MT network
to a ground state characterised by say a special configuration
of the displacement field $(j, m)$. This ground state
will be conformally invariant, and therefore a true vacuum
of the string, only after complete evaporation of the black hole,
which however would keep  memory of the particular collapse
mode in the `value' of the constant added to the dilaton field,
or other $W$-charges.
This
reflects the existence of additional exactly marginal
deformations, consisting of global modes only,
that are not direclty accessible by local scattering experiments,
in the context of the low energy theory of propagating modes
(displacement field configurations $\psi $).
In such a case, the resulting ground state will be infinitely
degenerate, which would solve the problem of {\it memory
capacity}
\pr
Breaking of this degeneracy, can be achieved
by means of an
{\it external stimulus}, which is believed to be
due to a weak field~\cite{delgiud}. For instance,
following the suggestion of ref. \cite{delgiud}, we
may imagine that an external weak field produces a
spontaneous breaking of the electric dipole rotational
symmetry in the water molecules, resulting in a
`lasering'~\cite{prep}
of the environmental surroundings of the MT system
(excitation of coherent dipole quanta).
Such an excitation of collective modes results
in a specific {\it code} characterising the ground state,
as we mentioned above.
The so selected
ground state of the ordered water molecules affects the
MT chains, due to the friction coupling $\rho $ in (\ref{ten}),
(\ref{dilaton}). As becomes clear from the analysis
above ((\ref{ten})-(\ref{13})),
the effects of the environment are described by selecting
a specific value for the vacuum expectation value (condensate)
of the dilaton field in our suggested stringy approach to MT
dynamics.
The other $W$-charges (moduli)
may also be selected this way,
which we believe corresponds to the {\it memory printing}
process, i.e. storage of information by a selection
of a given ground state. A new information would then
choose a different value of the dilaton field
or other $W$-hair charges,
etc. This provides a new and satisfactory mechanism of {\it memory
recall} in the following sense: if a new pulse happens to
correspond to the same set of (conserved)
$W$-hair moduli
configurations~\cite{emn}, then the associated virtual black hole
will be characterised by the same set of quantum hair, and
then the same memory state is reached {\it asymptotically}
(process of `memory
recall'). The discussion we gave in the previous section
about the r\^ole of
world-sheet instanton deformations in shifitng the
$ADM$ mass of a black hole
(\ref{adm}), while keeping memory of the dilaton v.e.v.,
finds a natural application in this coding
process. Moreover,
the irreversible arrow of time, endemic in Liouville
string theory~\cite{emn} explains naturally why ``only
the past can be recalled''~\cite{vitiello}.
\pr
To understand why the above process leads to a special coding,
and how time reversal is spontaneously broken, as a result
of this coding, it is sufficient to recall our
discussion
above, according to which in
the presence of a space-time foamy environment,
characterised by the virtual appearence and evaporation of black holes,
there is a coupling of global modes to the propagating modes.
As a result of the exactly marginal character of the
deformations~\cite{emn,chaudh},
which thus respect conformal invariance at a string level,
the environmental global
modes match in a special way with the propagating mode  $j=-\frac{1}{2}$
$m=0$, which is the
zero mode of the (massles)
tachyon corresponding to the tachyon background
of a two dimensional black hole which constitutes the {\it ground}
state or {\it memory} state of our system.
This is a {\it special
coding} which were it not for the infinte degeneracy
of the black hole space time would lead to
a restricted memory capacity\footnote{It should
be noted, once more, that the various other
(infinite) states corresponding to continuous representations
of the $SL(2,R)$ symmetry that pertain to various
tachyon modes do not constitute memory states,
because, as mentioned earlier,
they are just {\it excitations} about the ground state.}.
\pr
We cannot resist in
pointing out that the
existence of such coded situations in brain cells
bears an interesting resemblence with DNA coding,
with the important difference, however, that
here it occurs in the model's state space.
In this context, we note that the genetic diversity
is not due to an infinite number of nucleotide
types, since in nature the relevant ones are only four,
paired by two ($A~=~T/G~\equiv~C$), but
rather to a macroscopically
large number of existing combinations in the DNA helix.
Similarly, for the extremely rich (macroscopic) memory
capacity
in our stringy MT model, we may not need
the full
(infinite) set of the $W$-hair charges,
but just the
dilaton vev $<\Phi >$ may be sufficient as a `collective'
mode. The latter is, as we mentioned earlier, related
to external stimuli through the equations (\ref{ten})-(\ref{13}).
\pr
A final comment we would like to make, concerns
quantum uncertainties in the length of MT chains. 
In ref. \cite{mn} a prediction has been made, based on the 
stochasticity of the Liouville string theory~\cite{emn},
about quantum jumps of the tips of MT chains, as a 
result of quantum uncertainties.
In a subsequent article \cite{aemn} we have studied the 
propagation of light signals in a Liouville gravity
environment, and argued that there are induced 
bounds in measuring distances, $L$,  as a result of 
quantum-mechanical uncertainties, which 
behave like 
\be 
{\rm min}\{\delta L\} \sim  \sqrt{L L_s}
\label{aemnl}
\ee
where $L_s$ is a characteristic string length (minimum distance).
$\delta L$ is a measurability bound rather than an uncertainty, 
i.e. it is a quantity that characterizes the matter 
subsystem in interaction with a measuring apparatus.
Thus, the bounds (\ref{aemnl}) are appropriate in 
experimental tests of the above theory, where one tries to 
measure fluctuations in the tips of the 
MT chains. 
Eq. (\ref{aemnl}) is a consequence of a modified dispersion 
relation form massless photons as a result of the interaction 
with quantum gravitational degrees of freedom, which in the present 
model of MT are the $W_\infty$ global string modes. 
Such fluctuations are, therefore, length dependent, and 
constitute an interesting prediction for possible experimental 
tests.  
\pr
The above picture of casting 
the complicated MT dynamics into a simple completely integrable
$1+1$-dimensional string model in an effectively curved space time
(representing the formaltion of singularities as a result of external 
stimuli in the brain) opens exciting possibilities 
for a rigorous study of the formation of such singluarities.
From a $1+1$-dimensional space-time point of view
this problem is equivalent to the strongly-coupled 
problem of black hole formation in two-dimensional 
string theory. The problem seems complicated, given 
its strongly-coupled nature. However, recently a new 
way of treating such kind of problems became possible
through the use of generalized duality symmetries,
which map the strongly-coupled problem of strings propagating 
in singular space-time backgrounds into weakly coupled membrane theories,
which appear solvable. 
The membranes used in such an approach are characterized by 
the incorporation of open strings in their spectrum, with 
special world-sheet boundary conditions (of Dirichlet type)
for a subset of the space-time coordinates of the string.
From a critical string point of view this subset 
characterizes the collective coordinates of the 
solitonic string  background. 
\pr
It is therefore natural to treat our MT model in this way,
with the hope of getting more quantitative information 
about the formation of the singularities, the recoil 
of matter due to the fluctuations of space time etc. 
The two dimensional black-hole problem is a string soliton 
problem as discussed in ref. \cite{emn}. Moreover, 
in the present MT model, even the flat-space time limit 
is a kink, characterized by a certain set of collective coordinates.
In the past year such compicated string soliton $\sigma$-models 
can be described in an easier way if one employs open strings, 
i.e. $\sigma$ models on 
world-sheets with bounaries, and imposes appropriate boundary 
conditions for the 
target-space collective coordinates of the soliton. 
From a world-sheet point of view a target two-dimensional 
black hole model represents a defect on the world-sheet~\cite{emn,wbreak}
and therefore naturaly incorporates open strings in its spectrum.
The end points of such strings carry the gauge charges of the 
black-hole $W_\infty$-hair carrying information~\cite{emndbrane}. 
In this way one has an explicit (in a mathematical sense) way 
of describing storage of this information on the horizon of the 
singularity. Details have been given in ref. \cite{emndbrane},
where we refer the interested reader. From a brain MT 
view-point, 
the various values of such charges 
characterize the ground {\it memory} states, according to 
our discussion above~\cite{mn}. 
The presence of open strings carrying gauge group quantum 
numbers can also be seen in a totally different way
as follows: the two-dimensional black hole space time
has a cigar like structure, but it is symmetric about the 
spatial origin, i.e it has two asymptotically flat domains.
For consistency, one should get rid of one asymptotic
domain by some sort of orbifold procedure, i.e. appropriately
projecting the target space time of the string 
onto a single cigar. This induces, according to the analysis 
of \cite{horava}, open strings, and leads to an alternative
way of representing the two-dimensional black hole as a 
Dirichlet brane. 
\pr
Since this meeting, 
enormous 
progress has been made in understanding
the quantum mechanics of such Dirichlet branes~\cite{dirichlet}. 
In our group,
we have also made progress
towards a formal understanding of several issues 
arosen here, especially those 
regarding the connection of microscopic black-hole 
formation and evaporation with the  
above-mentioned revolutionary formalism of membrane theories. 
In particular, 
a special   
formalism on the world-sheet of the string 
has started to be developing 
with the hope of describing  interaction of matter with 
the membrane/black-hole
as accurately as possible. 
The approach utilizes certain 
`logarithmic operators'~\cite{log,diffusion}, which are special 
marginal world-sheet operators, 
capable of describing in a conformal-field-theory
setting changes of state of the membrane background. 
We have actually shown that
recoil of the (Dirichlet) string membrane 
world-volume during scattering 
of light matter off the membrane induces non-criticality of the 
pertinent ($\sigma$-model) 
string theory, associated with a change of 
state of the membrane background. This change of state
is described by the logarithmic operators. 
As discussed in ref. \cite{diffusion}, 
this results in decoherence 
for the effective subsystem of the light degrees of freedom,  
due to information carried by  
the (quantum) recoil degrees of freedom pertaining 
to fluctuations of the 
collective coordinates of the soliton~\cite{log}. 
Due to this decoherence, there are modified 
measurability bounds in lengths (\ref{aemnl}), 
implying modified 
(length-dependent ) uncertainty
relations 
in the non-crtitical string framework.
We believe, that this formal 
approach is useful towards a better  
understanding of important quantum aspects of black hole physics,
which have been argued in this talk and in refs. \cite{ideas,mn} 
to play a r\^ole in brain functioning. The recoil 
phenomenon, for instance, may be identified with a back-reaction
of the spin chain of the MT against an externally
induced abrupt change in the respective conformations
({\it stimulus}). We do hope to come back to such issues 
in more detail in the near future.  
\pr
\nk {\Large {\bf Acknowledgements}}
\pr
The work of D.V.N. is supported
in part by D.O.E. Grant
DEFG05-91-GR-40633.

\end{document}